\documentclass[aps,prl,twocolumn,floatfix]{revtex4-2}
\usepackage[utf8]{inputenc}
\usepackage[T1]{fontenc}
\usepackage[english]{babel}
\usepackage{mathtools}
\usepackage[scaled=1.005]{newtxtext}
\usepackage{amsmath}
\usepackage{amsfonts}
\usepackage[scaled=1.005,varbb,varg]{newtxmath} 
\usepackage{bm,dsfont}
\AtBeginDocument{\renewcommand{\vec}[1]{\bm{#1}}}
\newcommand{\pdagger}{\phantom{\dagger}}

\usepackage[dvipsnames,svgnames,x11names,hyperref]{xcolor}
\usepackage{hyperref}
\hypersetup{ 
    colorlinks=true, linkcolor=NavyBlue, urlcolor=NavyBlue, citecolor=NavyBlue
}

\usepackage{physics,braket,siunitx,slashed}
\usepackage[capitalise]{cleveref}

\begin{document}

\title{Spin and Orbital Metallic Magnetism in Rhombohedral Trilayer Graphene} 

\author{Chunli Huang}
\affiliation{Theoretical Division, T-4, Los Alamos National Laboratory, Los Alamos, New Mexico 87545, USA}
\author{Tobias M.~R.~Wolf}
\author{Wei Qin}
\author{Nemin Wei}
\author{Igor V. Blinov}
\author{Allan H.~MacDonald}
\affiliation{Department of Physics, University of Texas at Austin, Austin TX 78712}

\date{\today} 

\begin{abstract}
We provide a complete theoretical interpretation of the metallic broken spin/valley symmetry states recently discovered in ABC trilayer graphene (ABC) perturbed by a large transverse displacement field. 
Our conclusions combine insights from ABC trilayer graphene electronic structure models and mean field theory, and are guided by precise magneto-oscillation Fermi-surface-area measurements. 
We conclude that the physics of ABC trilayer graphene is shaped by the principle of momentum-space condensation, which favors Fermi surface reconstructions enabled by broken spin/valley flavor symmetries when the single-particle bands imply thin annular Fermi seas. 
We find one large outer Fermi surface enclosed majority-flavor states and one or more small inner hole-like Fermi surfaces enclosed minority-flavor states that are primarily responsible for nematic order.  
The smaller surfaces can rotate along a ring of van-Hove singularities or reconstruct into multiple Fermi surfaces with little cost in energy. 
We propose that the latter property is responsible for the quantum oscillation frequency fractionalization seen experimentally in some regions of the carrier-density/displacement-field phase diagram.
\end{abstract}

\maketitle

\begin{figure}[ht]
\centering
\includegraphics[width=1\columnwidth]{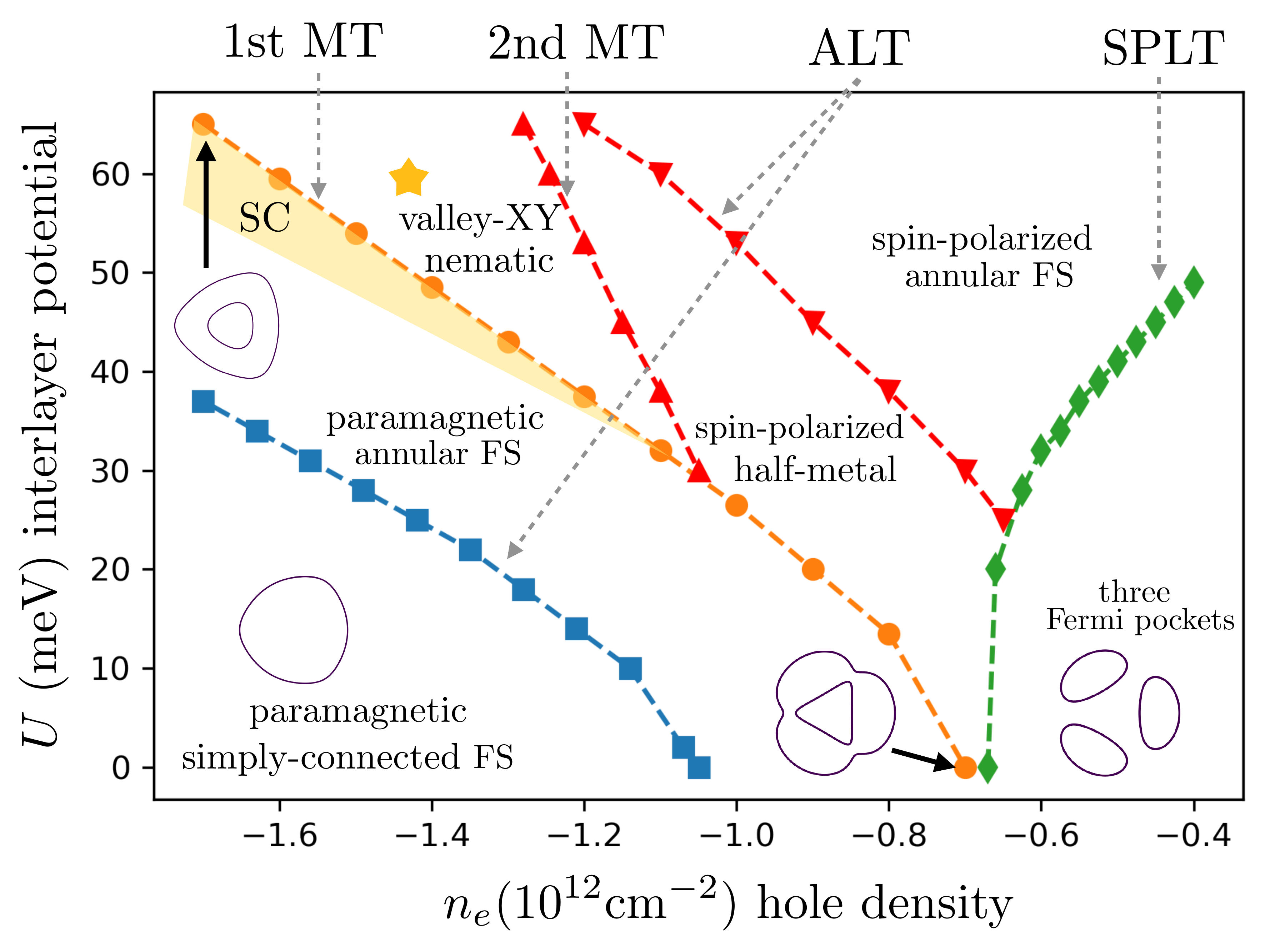}
\caption{%
Zero temperature phase diagram of ABC trilayer graphene.  
Fermi surface topologies in the various phases are indicated schematically.  
The phase boundaries mark magnetic transitions (MT) and Lifshitz transitions, both of which change Fermi surface topology. 
We distinguish two types of Lifshitz transitions, ALTs (annular Lifshitz transitions) that occur when the Fermi level crosses $\vec{k}=0$ energy band maxima to form an electron pocket inside an annular hole Fermi sea, and SPLTs (saddle point Lifshitz transitions) that occur when the Fermi level crosses $\vec{k}\ne0$ energy band saddle points to break the annulus into three separate pockets. 
At the first MT the paramagnetic annular Fermi sea supports a pair of Shubnikov-de~Haas quantum oscillations with normalized frequencies  $\vec{f}_{\nu}^{*}=(0.06,0.31)$. 
The interlayer potential $U$ is proportional to the displacement field $D$ in the experiment. Superconductivity occurs in region of the phase diagram shaded yellow close to the high hole density first MT. 
We used the dielectric constant $\epsilon_r=4$.
}
\label{fig:fig1}
\end{figure}

\textit{Introduction:--}
Multilayer graphene systems continue to surprise researchers with unexpected states of matter. 
Zhou.~\textit{et.al} \cite{zhou_half_2021,zhou_isospin_2021,zhou_superconductivity_2021} have recently uncovered rich phase diagrams in both AB bilayer and rhombohedral ABC trilayer graphene containing half-metal, quarter-metal, partially-isopsin polarized (PIP) metal, spin-triplet superconductor, and spin-singlet superconductor states in a three dimensional parameter space spanned by the electric displacement field $D$, the carrier density $n_e$, and the magnetic field $B$, see also Ref.~\cite{seiler2021quantum,2021arXiv211013907D}. 
The origin of superconductivity has been studied in Refs.~\cite{chatterjee_inter-valley_2021,ghazaryan2021unconventional,2021arXiv210904669Y,2022PhRvB.105g5432C,fRG_wei2022}.
In this Letter, we explain the pattern of spontaneous symmetry breaking and Fermi surface reconstructions observed in the normal metal states of ABC trilayer graphene, and comment on what they say about electron-interaction physics in two-dimensional multilayer graphene electron gases.

\begin{figure*}[t]
\centering
\includegraphics[width=2.0\columnwidth]{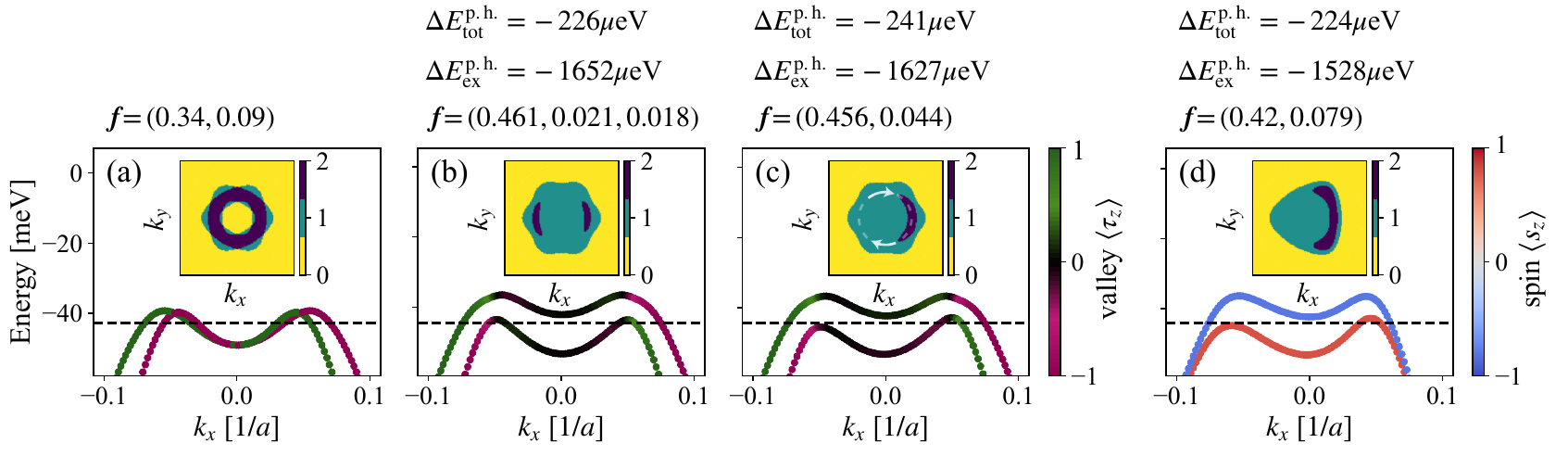}
\caption{%
Mean-field valence bands and Fermi surfaces of competing states at $n_e=-1.4\cdot 10^{12}\,\text{cm}^{-2}$ and $U=60$ meV (yellow star in Fig.~\ref{fig:fig1}): 
(a) the paramagnetic metal, (b--c) the valley-XY nematic metal and (d) the spin nematic metal.  
The yellow, green, and blue regions are respectively occupied by zero, one, and two holes. 
In the ordered region all magnetic states gain roughly the same amount of total energy relative to the paramagnetic state, ($\Delta E_{\text{tot}}^{\text{p.h.}}\sim 200\mu$eV for the illustrated case) by lowering exchange energy $\Delta E_{\text{ex}}^{\text{p.h.}}\sim 1500\mu$eV per hole at a cost in band energy. 
Their energy differences between the different ordered states ($\sim 1 \to 20\mu$eV per hole) is mainly determined by the characteristics of the inner nematic Fermi sea. 
The magnetic anisotropic energy in the $\mathrm{SU}(4)$ spin--valley space first minimizes the area of the inner hole Fermi sea, measured by the quantum oscillation frequencies $\vec{f}$, and then breaks rotational symmetry to make the inner surface more compact. 
The arrows in (c) suggest the sliding degree-of-freedom of the crescent-shaped Fermi sea. 
This mean-field calculation was performed with an accuracy of $\lesssim1 \mu$eV per hole, using $\epsilon_r=4.5$ and $D=50$nm.
}
\label{fig:fig2}
\end{figure*}

At low carrier densities $|n_e|\lesssim 2\times10^{12}\text{cm}^{-2}$ and large electric displacement fields $D\gtrsim 0.2\text{V/nm}$, holes ($n_e<0$) in ABC graphene occupy low velocity Bloch states \cite{zhang_band_2010,mccann_electronic_2013} with momenta near one of two inequivalent honeycomb lattice Brillouin-zone corners. 
These regions of momentum space are referred to below as valleys, and endow electrons with a valley pseudospin in addition to spin.  States in the two valleys are related by time-reversal symmetry. 
Because the spin- ($s=\uparrow,\downarrow$) and valley ($\tau=K,K'$) projected densities of states are identical for each spin/valley, the paramagnetic state occupies the four flavors equally. 
The broken spin/valley symmetry states seen experimentally occupy the four flavors unequally, just as magnetic metals occupy majority and minority spins unequally, and are expected because the Fermi energy at low carrier densities is small compared to the Coulomb energy per particle. 
Because of the high quality of the ABC graphene devices studied by Zhou {\it et al.}, it has been possible to measure magnetic oscillations at weak magnetic fields, and in this way to accurately measure the areas enclosed by most of the Fermi surfaces present in most phases. 
Magneto-oscillations do not on their own distinguish different symmetry-broken states and do not provide the information about Fermi surface shapes. 
These we infer by combining insights from ABC trilayer graphene electronic structure models and mean field theory calculations \cite{supmat} that are corroborated by features that appear in resistance {\it vs.~}$n_e$ and $D$ maps, in addition to the thermodynamic compressibility data \cite{zhou_half_2021,zhou_isospin_2021,zhou_superconductivity_2021}. 
We detail our band structure model and mean-field calculations in the Supplemental Material~\cite{supmat}.


We seek to explain the evolution of the ground state as the hole density decreases and interactions increase in importance. 
As shown in Fig.~\ref{fig:fig1}, at large interlayer potential $U$ the first transition is an annular Liftshitz transition (ALT) at which an electron pocket emerges in the interior of the hole Fermi sea.
This annular Fermi sea undergoes a first magnetic transition (1st MT) to a valley ordered state accompanied by a strong Fermi surface reconstruction; \textit{i.e.}, the number of Fermi surfaces (per spin) is reduced from four to two and the annulus near $k=0$ is removed. 
This state is susceptible to nematic broken symmetries that give rise to small crescent-shaped Fermi surfaces, which explain the small quantum oscillation frequencies reported in Ref.~\cite{zhou_half_2021,zhou_superconductivity_2021}.
We identify such a valley ordered state as a valley-XY nematic metal. \footnote{Note that other works sometimes refer to valley-XY ordered states as `intervalley coherent states`}. 
When the hole density is further lowered, the crescent-shaped (nematic) Fermi liquid shrinks and leads to a negative electronic compressibility as reported in Fig.~1d of Ref.~\cite{zhou_half_2021}. 
The crescent-shaped Fermi liquid disappears at a second magnetic transition (2nd MT) to a spin-polarized half-metal. 
As the hole densities is lowered even further, a second ALT changes the topology of the spin-polarized half-metal to an annular shape. 
Both the paramagnetic phase ALT (blue line) and the spin-polarized state ALT are prominent experimentally because they lead to easily identified peaks in resistance along the lines from $(n_e\,,\,D)=(-1.2,0)$ to $(-1.8,0.28)$ and from $(-0.6,0)$ to $(-1.1,0.5)$ in Fig.~1e of Ref.~\cite{zhou_half_2021},  respectively. 
At small $U$, there is only one magnetic transition because the paramagnetic metal avoids the transition into a valley-XY nematic metal. 
In what follows, we first address the instability of a paramagnetic metal with an annular Fermi sea \cite{jung2015persistent}, then the properties of magnetic metals.

\textit{Annular Fermi sea:--}
Depending on $n_e$ and $U$, the single-particle bands can have one of three distinct Fermi surface topologies (see Fig.~\ref{fig:fig1}): 
a simply-connected Fermi sea, a Fermi sea with three disjoint pockets, and an annular Fermi sea. 
With decreasing hole density, the annular topology appears at the ALT when the Fermi level intersects a shallow local valence band minimum at $\vec{k}=0$, and persists to the saddle-point Lifshitz transition (SPLT) at which the annulus splits into three pockets and the density-of-states diverges logarithmically.  
We focus below on the experimentally demonstrated instability of the $C_3$-distorted annular Fermi sea, which occurs well before the SPLT is reached and is therefore not simply because of the large density-of-states at the Fermi level.

The annular Fermi sea has an exterior hole-like Fermi surface and an interior electron-like Fermi surface, as shown in Fig.~\ref{fig:fig2}(a) \footnote{We overlay the Fermi surfaces centered at $K,K'$ point to the $\Gamma$ point for better comparison to the symmetry broken states.}.
The enclosed areas $A_{FS,\text{in}}$ and $A_{FS,\text{out}}$ lead to a pair of 
normalized quantum oscillation frequencies $\vec{f}_{\nu}=(f_{\nu,\text{in}}\,,\,f_{\nu,\text{out}})$:
\begin{equation}
f_{\nu,\text{in}}=\frac{ A_{FS,\text{in}} }{ \sum_{\alpha}  A_{h,\alpha}} \;  ,  \;
f_{\nu,\text{out}}=\frac{ A_{FS,\text{out}} }{ \sum_{\alpha}  A_{h,\alpha}},
\end{equation}
where the sum in the denominator is over the flavor index $\alpha=\{K\uparrow, K\downarrow, K'\uparrow,K'\downarrow\}$.
%
The paramagnetic metal partitions holes equally into the four flavors so that 
\begin{equation}  \label{eq:equal_area}
     A_{h,\alpha} \equiv A_h^0 = \frac{(2\pi)^2 n_e}{4} \; \; \forall\, \alpha. 
\end{equation}  
For an annular Fermi sea, $A_h^0=A_{FS,\text{out}}-A_{FS,\text{in}}$. 

\textit{Valley-XY nematic metal:--} 
As the inner Fermi surface of the paramagnet becomes larger, occupied hole momenta are more widely dispersed and this leads to a reduction in the long-range Coulomb exchange energy \cite{jung2015persistent}.
We identify this reduction as the driver of the first magnetic transition, which occurs near $A_{FS,\text{in}}/ A_h^0 \approx 0.3 $ and therefore with normalized quantum oscillation frequencies
\begin{equation}
\vec{f}_{\nu}^{*} \sim (0.06,0.31).
\end{equation}
The fact that these frequencies are nearly independent of $U$ emphasizes that Fermi sea shape is more important in driving the instability than hole density. 
%
%

The 1st MT leads to strong flavor polarization and Fermi surface reconstructions where the majority-flavor states have a large hole-like Fermi surface with no annulus. 
Since the outer Fermi surface $A_{FS,\text{out}}$ has a large Fermi velocity, it is energetically favorable to retain a small minority-flavor hole pocket centered at $k\neq0$ to limit its expansion. 
For a given $n_e$, the exchange energy favors the valley-XY state over the spin-polarized and valley-Ising states illustrated in Fig.~\ref{fig:fig2} because the outer Fermi surface of its valley-mixed quasiparticles has $C_6$ symmetry and can therefore enclose a larger area for given Fermi radii. 
The relationship between the spin--valley $\mathrm{SU}(4)$ magnetic anisotropy energy and the area of the inner Fermi surface is illustrated in Fig.~\ref{fig:fig2}(b,c) for valley-XY polarized states and in Fig.~\ref{fig:fig2}(d) for a spin-polarized state. 
The 1st MT leads to the discontinuous changes in quantum oscillation frequencies from the paramagnetic state values $\vec{f}_\nu^*$ to the ones shown in Fig.~\ref{fig:fig2}. 
Although the candidate ordered states differ little in energy, they do have distinct quantum oscillation frequencies.  
The values reported in Ref.~\cite{zhou_superconductivity_2021} support our identification of the valley-XY state as the ground state in this region of the phase diagram. 
Note that the quantum oscillation data put a severe constraint on the many possible magnetic states in the large $\mathrm{SU}(4)$ Hilbert space, see Ref.~\cite{supmat}. In particular, an intervalley coherent state (valley-XY) without nematicity cannot explain the sudden jump in the oscillation frequencies at the phase-boundary.



We now explain the crescent shape of the reconstructed inner Fermi sea of the valley-XY metal shown in Fig.~\ref{fig:fig2}(c). 
Note first that the valence bands have a ring of maxima at $\vec{k}_{\text{VBM}} \sim k_0(\cos(\theta),\sin(\theta))$ where $k_0a\sim0.05$ and $ 2\pi > \theta \geq 0$. 
When a point on the ring rises above the Fermi level it nucleates a hole-like Fermi sea. 
Such Fermi sea nucleation can be understood as a form of momentum-space condensation introduced by Heisenberg \cite{Heisenberg1984}. 
Because the energy dispersion along the radial direction is steep, due mainly to strong non-local exchange-splitting, whereas the dispersion along the ring of valence band maximum is anomalously flat, the nucleated Fermi sea has the form of a crescent. 
Furthermore, the crescent Fermi sea wants to be as compact as possible to increase the exchange energy from the long-range Coulomb interaction. 
Thus, although the reconstructed inner Fermi seas in Fig.~\ref{fig:fig2}(b,c) have the same area, their total energy differ because of the relative compactness of their Fermi seas. 
The single-crescent state is the most compact and hence has the lowest energy. 
The crescent Fermi sea leads to nematicity, which we characterize by the $2\times2$ traceless symmetric tensor
\begin{align}
    \hat{N}_{ij}= \hat{D}_{ij} -\frac{1}{2} \text{tr}(\hat{D}) \delta_{ij},
\end{align}
where $i,j\in\{x,y\}$ and $\hat{D}$ is the Drude tensor given by
\begin{align}
    \hat{D}_{ij}= \sum_{a=1,2...N_F}  \int  \left( \frac{v_{a,i}v_{a,j}}{|\vec{v}_a|} \right) \frac{dk_{\parallel,a}}{4\pi^2\hbar}.
\end{align}
Here $\vec{v}_a(\vec{k})=\nabla \epsilon_a(\vec{k})$ is the Fermi velocity and $a=1,2,\dots,N_F$ enumerates the Fermi surfaces. 
The eigenvectors of $\hat{N}$ are the directors of the nematic Fermi sea, and the eigenvalues $\pm\sqrt{\text{det}(\hat{N})}$ defines a nematic order parameter
\begin{equation}
    \mathcal{N} = \sqrt{|\text{det}(\hat{D})|}/\text{tr}(\hat{D}).
\end{equation}
The values for the single crescent and double crescent states are $\mathcal{N}=4\%$ and $\mathcal{N}=6\%$, respectively. 
This nematicity can lead to significant transport anisotropy below the transition temperature, that we predict will be most prominent close to the phase boundary of the 1st MT.

The crescent Fermi sea is very fragile in the sense that it costs very little energy to deform and slide it around the ring of valence band maxima at $\vec{k}_{\text{VBM}}$ as indicated in Fig.~\ref{fig:fig2}(c). 
We can control the center of mass of the crescent Fermi sea on $\vec{k}_{\text{VBM}}$ by minimising $\tilde{H}=H^{\text{MF}}- \mu_{\text{nem}} \, \rho(\vec{k}_{\text{VBM}})$ where $\mu_{\text{nem}} $ is a Lagrange multiplier, $H^{\text{MF}}$ is the mean-field Hamiltonian, and $\rho(\vec{k}_{\text{VBM}})$ is the density matrix evaluated at $\vec{k}_{\text{VBM}}$ \cite{supmat}. 
The pinning potential which favors particular crescent orientations is $\lesssim 1$ $\mu$eV per hole, and cannot be resolved by our finite $k$-grid mean-field calculations \footnote{We used a triangular $k$-grid with $140\times140$ point-resolution in regions up to radii $k_{\text{max}}a=0.2$}. 
When the crescent Fermi sea becomes small, it costs very little energy to split it into two or more smaller crescents. 
We attribute the fractionalization of the small quantum oscillation frequencies reported in Ref.~\cite{zhou_superconductivity_2021} as the system advances deeper into the valley-XY state, to this fragility. 

\begin{figure}
\centering
\includegraphics[width=1\columnwidth]{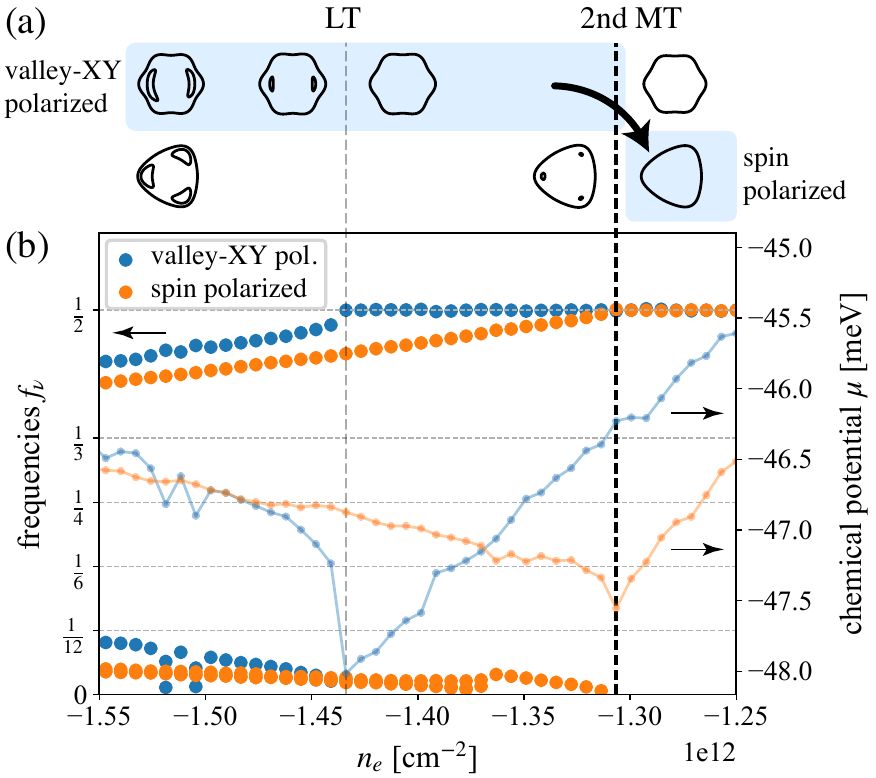}
\caption{%
Transition from a nematic valley-XY phase to the valley-XY half-metal and to the spin-polarized half-metal as a function of decreasing hole density $n_e$ at fixed interlayer potential $U=62$~meV. 
(a) The evolution of the Fermi sea for valley-XY and spin ordered states, where the ground states are shaded light blue. 
The valley-XY ordered state undergoes a Lifshitz transition (LT) upon which it loses its inner pockets before going through the 2nd magnetic transition (MT). 
(b) Quantum oscillation frequencies $f_\nu$ (big markers) and chemical potential $\mu$ as the electron density increases.
This calculation was performed with model parameters $\epsilon_r=4.0$, $d_{\text{gate}}=50$~nm, and a $k$-mesh with $N_k=10^4$ points.
}
\label{fig:fig3}
\end{figure}

\textit{2nd magnetic transition:--} 
As the hole density decreases in the valley-XY metal phase, the paramagnetic region enclosed by the crescent Fermi sea shrinks as shown in Fig.~\ref{fig:fig3}(a) and increases the valley-XY order parameter. 
This leads a decrease of Fermi energy ($\mu$) as shown in Fig.~\ref{fig:fig3}(b) and explains the observed (slightly) negative inverse compressibility $\kappa=\partial \mu/\partial n_{e}$ in the PIP phase of Ref.~\cite{zhou_half_2021}.
When the nematic Fermi sea disappears completely, $\mathcal{N}=0$ and $\vec{f}_{\nu}=(0,0.5)$. 
The valley-XY half-metal is then characterized by a single compact Fermi surface with $C_6$ rotational symmetry (for each spin). 
Although the $C_6$ Fermi surface is closer to a circle, compared to the spin-polarized half-metal which only has $C_3$, the content of the wavefunction in the valley-XY metal changes dramatically over the enclosed area. 
The valley-XY half-metal eventually undergoes a 2nd magnetic transition to a spin-polarized half-metal, as shown in Fig.~\ref{fig:fig1} and indicated by the arrow in Fig.~\ref{fig:fig3}(a).
Although this is a first-order phase transition, there is no change in $\vec{f}$ because both states are half-metals, and no change in $\mathcal{N}$ because both states lack nematicity. 
Their phase boundary, however, can easily be shifted by an in-plane spin-polarizing Zeeman field \cite{zhou_half_2021,zhou_isospin_2021,zhou_superconductivity_2021}. 
The transition to the spin-polarized half-metal can be understood as a reset transition \cite{zhou_half_2021,zondiner2020cascade,saito2021isospin,wong2020cascade}, at which the minority spin Hilbert space becomes inactive while there is still equipartition of density in the majority spin Hilbert space: 
\begin{equation}
    A_{h,K\uparrow}= A_{h,K'\uparrow}= \frac{(2\pi)^2n_e}{2} \;\; ,\;\; A_{h,K\downarrow}= A_{h,K'\downarrow}=0.
\end{equation}
As the hole density is lowered, a Lifshitz transition creates an annular topology in the (majority) $\uparrow$-spin Fermi sea and the same symmetry-breaking pattern repeats in the $\uparrow$-spin Hilbert space as the density of holes is further lowered.

\textit{Spin-fluctuation and spin-orbit interactions:--}
At weaker displacement fields, the Fermi velocity of the outer Fermi surface is smaller at the first magnetic phase transition than it is at large $U,|n_e|$, especially where it is close to the three SPLT momenta. 
It is hence favorable to make a direct transition from the paramagnetic state to the spin-polarized half-metal \footnote{The groundstate energy of a spin-polarized half metal and a valley-polarized half-metal appear to be almost degenerate due to the $SU(4)$ invariant Coulomb interaction}. 
Proximity to a transition between a paramagnetic metal and a ferromagnetic metal is known to suppress spin singlet superconductivity \cite{berk1966effect} and valley-singlet superconductivity \cite{huang2021pseudospin}, explaining the termination of singlet superconductivity at small $D$ seen in Ref.~\cite{zhou_superconductivity_2021}, which occurs despite an increasing density-of-states.
The direct transition from the paramagnetic state to the spin-polarized half metal leads to a sudden jump of quantum oscillation frequencies directly from $\vec{f}_{\nu}^{*}$ to $\vec{f}_{\nu}=(0,0.5)$. 
In our mean-field calculations, this direct transition continues to zero displacement field ($U=0$), as shown in Fig.~\ref{fig:fig1}. 
In experiment, however, the magnetic phase fades away at small displacement field and blends into the paramagnetic phase.  
We attribute this behavior to spin fluctuations that become important when the electric displacement field is small. 
The displacement field can suppress spin fluctuations because it increases the strength of spin-orbit coupling \cite{konschuh2012theory,huertas2006spin,min2006intrinsic}, providing the magnetic anisotropy needed for spin-order.

 \textit{Summary and Discussions--}
The magnetic order recently discovered \cite{zhou_half_2021,zhou_isospin_2021,zhou_superconductivity_2021} in metallic ABC trilayer graphene is unusual because it involves valley as well as spin degrees of freedom, and because the quasiparticle bands imply that the unordered paramagnetic states have annular Fermi seas. 
By combining mean-field theory with a model that captures pertinent electronic structure details, we are able to account for the phase transitions observed experimentally and shed light on the nature of the different phases.  
In particular we identify the experimental partially-isospin-polarized (PIP) state as a valley-XY nematic metal. 
We find that, as in conventional itinerant electron magnets, the magnetic condensation energy greatly exceeds a much smaller magnetic anisotropy energy.  
The $\mathrm{SU}(4)$ spin--valley magnetic anisotropy energy controls a competition between valley and spin-polarized states, which are very close in energy throughout the phase diagram. 
We predict that the valley-XY metal, which is stabilized at large interlayer potential, is nematic because small minority-hole pockets, driven by the energetic expense of expanding the outer Fermi surface, break rotational symmetries in order to become more compact.  
The nematicity of these states should be observable in transport measurements. When the minority-holes are absent at low hole densities and at low displacement fields, spin order is preferred over valley order. In closing, let us briefly mention the possibility for the small nematic Fermi sea to crystallize into a Wigner crystal state that breaks both translation and $C_3$ symmetry. 
In the spirit of Ref.~\cite{berg2012liquidcrystal}, at sufficiently low density of holes, such anisotropic Wigner crystal states can have lower energy than the nematic fluid and conventional hexagonal Wigner crystal states due to the ring of Van Hove singularities in k-space. 

\textit{Note added:} 
Recently, we found that Ref.~\cite{2021arXiv211015254D} discusses momentum condensation (``flocking'' effect) in Bernal bilayer graphene with three disjoint Fermi pockets whose microscopic origin is similar to the formation of nematic Fermi surface we discuss here in ABC trilayer graphene. 
 
\begin{acknowledgments}
\textit{Acknowledgments:--} 
We thank Anna Seiler, Andrea Young and Haoxin Zhou for informative discussions.  
The work done at LANL was carried out under the auspices of the US DOE NNSA under Contract No.~89233218CNA000001 through the LDRD Program. 
This work was supported by the U.S. Department of Energy, Office of Science, Basic Energy Sciences, under Award DE-SC0022106. 
T.M.R.W.~is grateful for the financial support from the Swiss National Science Foundation (Postdoc.Mobility Grant No.~203152).
\end{acknowledgments}

\bibliography{references}

\newpage
\widetext
\begin{center}
\textbf{\large Supplementary Material: Pseudospin Paramagnons and the Superconducting Dome in Magic Angle Twisted Bilayer Graphene}
\end{center}
\setcounter{equation}{0}
\setcounter{figure}{0}
\setcounter{table}{0}
\setcounter{page}{1}
\makeatletter
\renewcommand{\theequation}{S\arabic{equation}}
\renewcommand{\thefigure}{S\arabic{figure}}

\title{Supplemental Material:\\ Spin and Orbital Metallic Magnetism in Rhombohedral Trilayer Graphene}

\section{Band structure model}

Rhombohedral trilayer graphene (RTG) has three layers $\ell=1,2,3$ separated by interlayer distance $d=3.4$~{\AA} and two lattices $\sigma=A,B$ each, such that the unit cell of the triangular lattice with lattice constant of $a=2.46$ {\AA} contains $6$ atoms. The layers of RTG are in a ABC stacking configuration.
Following Refs.~\cite{zhang_band_2010,koshino_trigonal_2009}, we consider a single electronic orbital ($\pi$) per atom, with spin $s=\pm 1/2$, and use the continuum model that can be derived from a tight-binding Hamiltonian by expanding the low-energy dispersion around each valley $\tau K$ with $\tau=\pm 1$ at the corners of the Brillouin zone. We label the four flavor combinations of spin $s$ and valley $\tau$ with the flavor index $\alpha$. In the basis $\psi_{\alpha}(\vec{k})=(\psi_{\alpha A1}(\vec{k}), \psi_{\alpha B1}(\vec{k}), \psi_{\alpha A2}(\vec{k}), \psi_{\alpha B2}(\vec{k}), \psi_{\alpha A3}(\vec{k}), \psi_{\alpha B3}(\vec{k}))$, the resulting continuum Hamiltonian for given spin--valley $\alpha$  is \cite{zhang_band_2010}
\begin{align}\label{eq:continuum_hamiltonian}
    h(\vec{k}) = \begin{bmatrix}
    t(\vec{k}) + U_1 & t_{12}(\vec{k}) & t_{13} \\
    t_{12}^\dagger(\vec{k}) & t(\vec{k}) + U_2 & t_{12}(\vec{k}) \\
    t_{13}^\dagger & t_{12}^\dagger(\vec{k}) & t(\vec{k}) + U_3
    \end{bmatrix}_{6\times 6},
\end{align}
where $\vec{k}=(k_x,k_y)$ is the Bloch momentum measured with respect to the center of valley $\tau$. This Hamiltonian contains matrices with intralayer hopping amplitudes ($t$), nearest-layer hopping amplitudes ($t_{12}$), and next-nearest-layer hopping amplitudes ($t_{13}$), as well as possible potential differences between different layers and/or sublattices due to external gates and/or broken symmetries (terms $U_i$). Explicitly, the various hopping amplitudes are given by
\begin{align}
    t(\vec{k}) = \begin{bmatrix}
      0 & v_0 \pi^\dagger \\
      v_0 \pi & 0
    \end{bmatrix}, \quad 
    t_{12}(\vec{k}) = \begin{bmatrix}
      -v_4 \pi^\dagger & v_3 \pi \\
      \gamma_1 & -v_4 \pi^\dagger
    \end{bmatrix}, \quad
    t_{13}(\vec{k}) = \begin{bmatrix}
      0 & \gamma_2/2 \\
      0 & 0
    \end{bmatrix},
\end{align}
where $\pi = \tau k_x+i k_y$ is a linear momentum, and $\gamma_i$ ($i=0,\dots, 4$) are hopping amplitudes of the tight-binding model with corresponding velocity parameters $v_i=(\sqrt{3}/2)a\gamma_i/\hbar$. We use the model parameters listed in \cref{tab:graphene_params_ABC}, which are chosen to match quantum oscillation frequency signatures in Ref.~\cite{zhou_half_2021}. 

\begin{table}[b]
\caption{\label{tab:graphene_params_ABC}Tight-binding parameters (in eV) for rhombohedral trilayer graphene, see also Refs.~\cite{zhang_band_2010,koshino_trigonal_2009,zhou_half_2021}.}

\begin{ruledtabular}
\begin{tabular}{llllllll}
$\gamma_0$ & $\gamma_1$ & $\gamma_2$ & $\gamma_3$ & $\gamma_4$ & $U$ & $\Delta$ & $\delta$ \\
$3.160$ & $0.380$ & $-0.015$ & $-0.290$ & $0.141$ & $0.030$ & $-0.0023$ & $-0.0105$ \\
\end{tabular} 
\end{ruledtabular} 
\end{table}

\begin{figure}[h]
\centering
\includegraphics[width=1.0\textwidth]{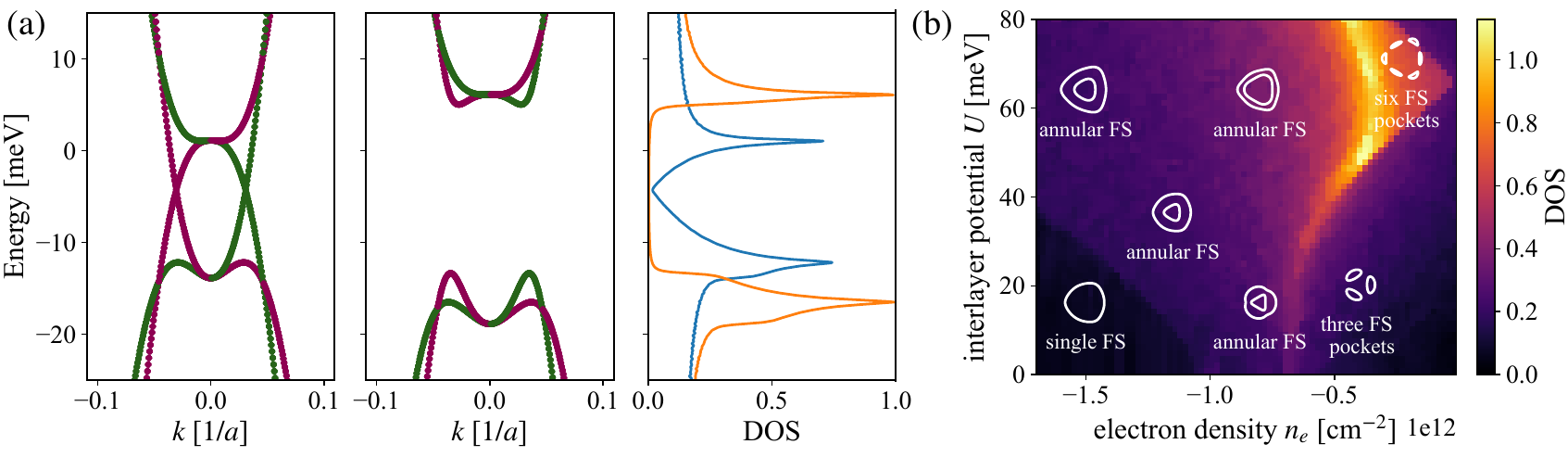}
\caption{Non-interacting band structure of rhombohedral trilayer graphene. (a) Electronic bands for valley $K$ (purple) and valley $K'$ (green) without and with interlayer potential $U=20$ meV, and corresponding density of states. (b) Density of states (DOS) at the Fermi surface as function of electronic density $n_e$ and interlayer potential $U$. Insets indicate the Fermi sea (FS) shape and topology per valley--spin found in different regions of the parameter space. The Lifshitz transition from single FS to annular FS is accompanied by a jump in the DOS, while transitions from annular to disjoint pockets is accompanied by a log-divergent saddle-point van-Hove singularity.}
\label{fig:noninteracting_bands}
\end{figure}

As shown in \cref{fig:noninteracting_bands}(a), the electronic band structure of $h^{\alpha}(\vec{k})$ in \cref{eq:continuum_hamiltonian} is strongly trigonally-warped (by interlayer hopping amplitudes) and has van Hove singularities near charge neutrality. The dispersion is semimetalic in absence of a interlayer potential $U$ but has a insulating gap when it is finite. In our work, we focus on the hole-doped side of the dispersion, in the electronic density range of $n_e \in[-1.8\cdot 10^{12},0]\;\text{cm}^{-2}$ accessible to recent experiments \cite{zhou_half_2021}, and consider interlayer potentials in the range $U\in [0,80]$ meV, corresponding to displacement fields $D\in[0,0.5]\; \text{V}/\text{nm}$ after including Hartree screening between the system and gates \cite{mccann_electronic_2013}. The latter is done implicitly by matching the quantum oscillation frequency data \cite{zhou_half_2021} at finite interlayer potential to band structure calculations. 
As shown in Fig.~\cref{fig:noninteracting_bands}(b), the topology and shape of the Fermi sea (FS) undergoes Lifshitz transitions when the electron density $n_e$ and the potentials $U$ changes over the experimentally accessible range: from a single simple FS to an annular FS, and then to three or six FS pockets. Each transition is accompanied by a van-Hove singularity that induces a discontinuity in the density of states: a jump in the case of going from single FS to annular FS, and a  saddle-point log-divergence when tuning towards isolated pockets.

\section{Coulomb interactions}

\paragraph{Hamiltonian} In this work, we investigate the ground state of $H=H_0+H_{\rm ee}$, where $H_0$ is the electronic Hamiltonian of rhombohedral trilayer graphene as in \cref{eq:continuum_hamiltonian} and $H_{\rm ee}$ is the Coulomb interaction between electrons. 
In second quantization, they are
\begin{align}\label{eq:hamiltonian_interaction}
    H = H_0+H_{\rm ee} &= \sum_{\vec{k}, \alpha, ij} \psi_{\alpha i}^\dagger(\vec{k}) \, h_{ij}(\tau k_x, k_y) \, \psi^{\pdagger}_{\alpha j} (\vec{k}) +  \frac{1}{2A}\sum_{\vec{k}, \vec{k}'\!\!, \vec{q}} \! V_{\vec{q}} \,
    \psi_{\alpha i}^\dagger(\vec{k}\!+\!\vec{q}) \psi_{\beta j}^\dagger(\vec{k}'\!\!-\!\vec{q})
    \psi^{\pdagger}_{\beta j}(\vec{k}')  \psi^{\pdagger}_{\alpha i}(\vec{k}),
\end{align}
where $\vec{k}=(k_x,k_y)$ is the same Bloch momentum as before, $i,j$ run through the six atoms in the unit cell, $\alpha,\beta$ run through the four flavor combinations of spin and valley, $A$ is the system area, and $V_{\vec{q}}$ is the electron--electron interaction potential in Fourier space. Note that the interaction here only depends on the in-plane distance between electrons, but is independent of layer, spin and valley degrees of freedom. The layer dependence is small and was neglected, which does not qualitatively affect our result. The spin--valley symmetry in our interaction is $\mathrm{SU}(4)$ since we neglect intervalley Coulomb scattering, which would reduce it to $\mathrm{SU}_{\text{spin}}(2)\times\mathrm{SU}_{\text{spin}}(2)$ (the same as that of the non-interacting Hamiltonian). 
As explained below, we use the dual-gate-screened interaction potential
\begin{align}\label{eq:coulomb_potential}
    V_{\vec{q}}=\frac{2\pi k_e}{\epsilon_r} \frac{\tanh(\vert\vec{q}\vert D)}{\vert\vec{q}\vert}
\end{align}
with metallic gates at distance $D \sim 30,\dots,50$~nm, relative permeability $\epsilon_r\sim 4$--$10$, and Coulomb constant $k_e=1.44$~eV~nm. The parameters $D$ and $\epsilon_r$ are treated as tuning parameters in our work, and are chosen informed by the onset of symmetry breaking observed in the experiment. Screening due to electron--electron interactions is not taken into account. We argue that the internal screening plays a smaller role than the gate screening, because the latter already removes the $\vec{q}\to 0$ divergence and exponentially cuts of the range in $k$-space of the interaction potential.

\paragraph{Gate screening} Gate screening is taken into account by considering the mirror charges induced by the metallic surface of nearby gates, which in the dual-gated case leads to an effective potential
\begin{align}
V(\vec{r},\vec{r}') = \frac{k_e}{\epsilon_r} \sum_{n=-\infty}^{\infty} (-1)^n \frac{1}{\sqrt{((\vec{r}_\perp - \vec{r}'_\perp)^2 + \xi_n^2}},
\end{align}
where $\vec{r}_\perp^{(\prime)}$ denotes the in-plane position and  $\xi_n = r_z - (-1)^n (r_z' - n D)$ is the vertical distance from a test charge to other (mirror) charges. The Fourier transform of this potential can be evaluated to
\begin{align}
V(q) &= \frac{2\pi k_e}{\epsilon_r q} \sum_{n=-\infty}^{\infty} (-1)^n e^{-\xi_n q}
 = \frac{2\pi k_e}{\epsilon_r q} 
 \begin{cases}
 	\frac{\cosh(D q) - \cosh(d q)}{\sinh(Dq)} & \text{intralayer} \\
   \frac{\cosh((D-d)q)-1}{\sinh(D q)} & \text{interlayer}
\end{cases}
\end{align}
where $q=\vert \vec{q}_\perp \vert$ is the in-plane momentum, and simplifies to \cref{eq:coulomb_potential} if $D\gg d$. 

\section{Self-consistent Hartree-Fock approximation}

We seek to minimize the the total energy $E_{\text{tot}}$ of $H=H_0+H_{\rm ee}$ within the subspace of Slater-determinant states. 
In this variational ansatz, we have $E_{\text{tot}}\approx\tr(\rho [h+h^{\text{SCMF}}(\rho)])/(2\tr(\rho))$ with a density matrix $\rho$ that satisfies the self-consistent equations
\begin{align}\label{eq:meanfield}
h^{\text{SCMF}}(\rho, \vec{k})\ket{n\vec{k}} = \varepsilon_{n\vec{k}} \ket{n\vec{k}}, 
\quad
\rho \!=\! \frac{1}{N_k}\sum_{n\vec{k}} f_{n\vec{k}} \ketbra{n\vec{k}}{n\vec{k}},
\end{align}
where $f_{n\vec{k}}$ are Fermi-Dirac weights and $h^{\text{SCMF}}(\rho)=h+\Sigma^{H}(\rho)+\Sigma^{F}(\rho)$ is an effective single-particle Hamiltonian.
The latter contains a Hartree potential and a Fock exchange term, i.e.,
\begin{align}\label{eq:hartree_selfenergyop}
\Sigma^{H}_{ab}(\rho) = \delta_{ab} \frac{1}{A} \sum_{b} V^{a b}(\vec{q}=0) \, n_b, &&
\Sigma^F_{ab}(\rho,\vec{k}) = - \frac{1}{A} \sum_{\vec{k}'} V_{ab}(\vec{k}-\vec{k}')\,\rho_{ab}^*(\vec{k}') ,
\end{align}
that contribute distinctly to $E_{\text{tot}}=E_{\text{band}}+E_{H}+E_{F}$.
Hartree describes how the total electronic density in degree of freedom $b$ affects (repels if $V$ is positive) electrons in degree of freedom $a$, and is just a classical electrostatic effect. The Fock term is a quantum mechanical contribution due to the antisymmetry of Fermions under exchange.
The chemical potential $\mu$ is constrained through the density $n_e A=N_k^{-1}\sum_{n\vec{k}} f_{n\vec{k}}$, where $N_k$ is the number of discrete k-points. 

Solving \cref{eq:meanfield} requires iterative methods that relax an initial guess $\rho_0$ into a fixpoint. 
We identify lowest-energy solutions by exploring different initial states $\rho_0$ (e.g., random, valley-polarized, spin-polarized ones). 
Substituting $h^{\text{SCMF}}$ with $h^{\text{SCMF}}(\rho)-\lambda \, X$ in \cref{eq:meanfield} imposes an additional soft constraint: $\lambda$ is a (small) Lagrange multiplier that favors the order parameter $\langle X \rangle$. In the main text, we used $X=\rho(\vec{k}_{\text{VBM}})$ to demonstrate how the crescent Fermi surface can be seeded at any point along the valence band maximum ring.

\section{Intervalley coherent state without nematicity}
\begin{figure}[h]
\centering
\includegraphics[width=12cm]{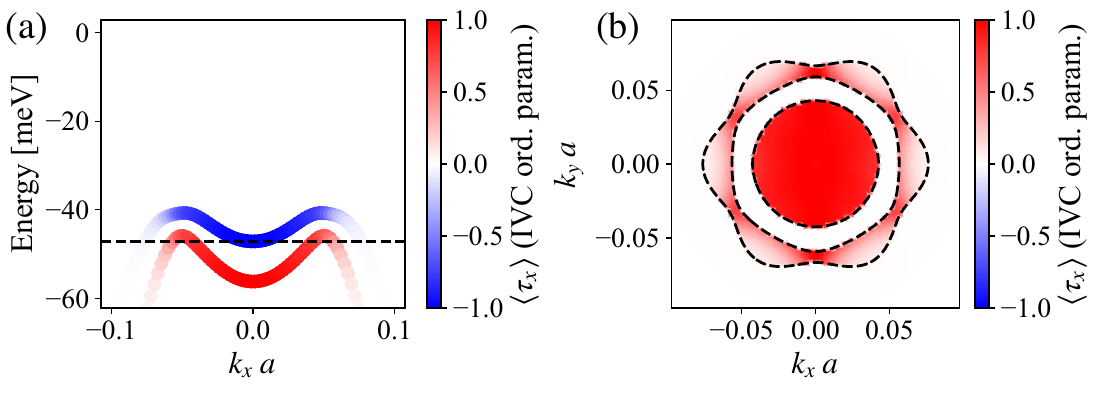}
\caption{A local minimum mean-field solution of the Hartree-Fock equation which shows finite intervalley coherence $\langle\tau_x\rangle\neq0$ but no nematicity.
(a) Intervalley-mixed quasiparticle bands and (b) corresponding Fermi sea where the color scale indicates the magnitude of the order parameter $\tau_x$.  The characters of the three Fermi surfaces with increasing radii are respectively electron-like, hole-like and hole-like. 
This state is calculated with dielectric permeability $\epsilon_r=4.3$, gate distance $D=50$nm at electron density $n_e=-1.75\cdot 10^{12} \text{cm}^{-2}$ and interlayer potential $U=70$meV. While this is a local minimum of the Hartree-Fock equation, the solution with finite nematicity we described in the main text has a lower energy.}
\label{fig:IVC_state_symmetric}
\end{figure}

In this section, we show that an intervalley coherent state without nematicity cannot explain the sudden jump of Shubnikov-de Haas oscillation frequencies at the first magnetic transitions.
In order to understand this, we consider an intervalley coherent state in \cref{fig:IVC_state_symmetric} which has two hole-like Fermi surfaces and one electron-like Fermi surface (the smallest one). The areas enclosed by these Fermi surfaces lead to the following three normalized frequencies 
   $ \vec{f}=(0.39, 0.25, 0.14)$.
 Note that the difference between the total area enclosing holes ($0.39+0.25=0.64$) and the area enclosing electrons ($0.14$) is half. 
However, such a state is not consistent with the experimental finding which has only two prominent frequencies $\vec{f}_{\rm exp}\sim(0.42,0.08)$ suggesting that there are only two hole-like Fermi surfaces \cite{zhou_superconductivity_2021}. The electron-like Fermi surface can be removed when electrons cluster at a point on the ring of van Hove singularities (valence band maxima) in \cref{fig:IVC_state_symmetric}(a) and this leads to the valley-XY nematic metal described in the main text.

\end{document}